\documentclass[10pt]{iopart}

\usepackage{iopams}  
\usepackage{graphicx}
\usepackage{amssymb}
\usepackage{epstopdf}
\usepackage{bbold}
\usepackage{float}
\usepackage{color}
\usepackage{subfigure}
\usepackage{cite}
\usepackage{relsize}
\expandafter\let\csname equation*\endcsname\relax

\expandafter\let\csname endequation*\endcsname\relax

\usepackage{mathtools}
\usepackage[ngerman, english]{babel}
\ioptwocol
\begin{document}

\title{Ring Rydberg Composites}
\author{Matthew T. Eiles$^*$, Andrew L. Hunter, Jan M. Rost}

\address{Max-Planck-Institut f\"ur Physik komplexer Systeme, N\"othnitzer Str.\ 38,
D-01187 Dresden, Germany }

\vspace{10pt}
\ead{\mailto{meiles@mpg.pks.de*}}
\date{\today}

\begin{abstract} The properties and behaviour of a Ring Rydberg Composite are explicated. This system consists of a ring of ground state atoms centered on a Rydberg atom, whose electron elastically scatters off the ground state atoms. We transform the electronic Hamiltonian into a discrete tight-binding representation in which the on-site energies and long-range hopping between sites are controlled and mediated by the Rydberg electron. From this new representation, which to a large extent enables an analytic treatment, we derive  scaling laws and analytic expressions for the wave functions and eigenspectrum.  The interface between ring and Rydberg geometries leads to a range of rich properties which can be tuned as a function of ring size, number of scatterers, and principal quantum number. 
\end{abstract}

\newcommand{\cev}[1]{\reflectbox{\ensuremath{\vec{\reflectbox{\ensuremath{#1}}}}}}

\newcommand{\be}{\begin{equation}}
\newcommand{\ee}{\end{equation}}
\newcommand{\dd}[1]{\mathrm{d}{#1}}
\newcommand{\ddd}[1]{\mathrm{d}^3{#1}}
\newcommand{\ddn}[2]{\mathrm{d^{#1}}{#2}}
\def\mbf{\mathbf}
\def\tbf{\textbf}
\def\red{\textcolor{red}}
\def\blue{\textcolor{blue}}
\def\JM{\textcolor{magenta}}
\def\i{\text{i}}
\newcommand{\bra}[1]{\langle{#1}|}
\newcommand{\ket}[1]{|{#1}\rangle}
\newcommand{\bkt}[2]{\langle{#1}|{#2}\rangle}

\section{Introduction} 
Since the early days of the quantum theory, the interaction of excited Rydberg atoms with ground state atoms has  provided crucial insight into various arenas of atomic physics, including plasma formation, collisional processes, and fundamental atomic properties \cite{Lebedev,LebedevFabrikant,GallagherPillet,GallagherBook,Fermi,Omont}. With the advent of ultracold Rydberg excitations in Bose-Einstein condensates, a  subfield devoted to the study of long-range Rydberg molecules, known colloquially as trilobite molecules, has developed \cite{Greene2000,KhuskivadzePRA,bendkowsky}. Experiments have resolved the vibrational spectra of Rydberg dimers, trimers, and so forth, and have sparked significant theoretical interest from the perspective of many-body physics \cite{WhalenPoly,Whalen2,Rost2006,Rost2009,quantumreflection,MolSpec,JPBdens,FeyKurz,FeyNew,FeyTrimer,EilesHyd,Schlag,EilesTutorial, RydbergRev,Demler}.   At the same time, attention has shifted towards experiments which use the Rydberg atom to probe the surrounding condensate and establish a ``microscopic laboratory'' to study  electron-atom and ion-atom scattering \cite{UltracoldChem,IonColdMeinert,IonRydBlockade, PhasesFeyMeinert,NewPfau,Sass,SpinFey,MacLennan}. 

Recently, we introduced the Rydberg Composite: a Rydberg atom with principle quantum number $\nu$ coupled to a large number of ground state atoms (scatterers) a well-defined regular or irregular arrangement, such as can be attained in a dense optical lattice or in a typical ultracold gas, respectively. The Rydberg Composite combines the high degeneracy, characteristic eigenfunctions, and scaling behavior of the Coulomb interaction with the engineered or structured environments typical of the solid state. Ref. \cite{PRXus} details the general structure of Rydberg Composites involving a one, two, and three-dimensional scatterer arrangements in a regular lattice. 

Here, we formulate the Ring Rydberg Composite, a ring of scatterers surrounding the Rydberg atom. This system has many desirable features, e.g. rich tunability and the possibility to address transport processes. Most importantly, we show how it can be described analytically to a large extent after formulating it using a tight-binding lattice Hamiltonian. This describes the hopping of a ``trilobite'' excitation across the ring sites, with on-site energies, dispersion relations, and inter-site interactions mediated by the electron and tunable by varying the ring size, number of scatterers, or level of Rydberg excitation. We focus on the characterization of the key features of this system and the development of analytic scaling behavior in order to build a foundation for future work utilizing the unique possibilities of this system. An experimental signature of immediate interest is the dependence of the ground state energy as a function of $M$.  

\section{Rydberg {C}omposites in the trilobite basis } 
In atomic units, the electronic dynamics of a Rydberg Composite is governed by the Hamiltonian 
\begin{equation}
\label{eq:RydbergbasisHamiltonian}
H _0= -\sum_{lm}\frac{\ket{\nu lm}\bra{\nu lm}}{2(\nu-\mu_l)^2}+2\pi \sum_{i=1}^Ma_s(R_i)\ket{\vec R_i}\bra{\vec R_i}.
\end{equation} 
The first term describes the bare Rydberg atom in terms of its eigenstates $\bkt{\vec r}{\nu lm} = \frac{u_{\nu l}(r)}{r}Y_{lm}(\hat r)$, where $m=-l,\dots,l$ and $l=0,\dots\nu-1$ are the magnetic and orbital angular momentum quantum numbers, respectively. The Hamiltonian is truncated to a single principal quantum number $\nu$, and in the following we neglect for simplicity the few non-zero quantum defects $\mu_l$ of an alkali atom. The second term describes the interaction between the Rydberg electron and $M$ scatterers placed at arbitrary positions $\vec R_i$, $i = 1,...,M$. This interaction, following the seminal work of Fermi,  is approximated by a contact potential parametrized by the $s$-wave scattering length $a_s$ \cite{Fermi}. The Hamiltonian of Eq. \ref{eq:RydbergbasisHamiltonian} neglects all other atom-atom and atom-ion interactions, and furthermore we assume a frozen gas scenario in which the scatterers do not move \cite{PRXus}.

Eq. \ref{eq:RydbergbasisHamiltonian} takes on a more appealing form when transformed into the so-called ``trilobite'' basis  \cite{JPBdens,Rost2006,Rost2009,FeyTrimer,EilesHyd,EilesTutorial}, whose states are the individual trilobite wave functions associated with each scatterer. The trilobite state $\ket{J}$ for a scatterer at position $\vec R_j$ is defined
\be
\label{eq:trilobasisdef}
\ket{J} = \sum_{l=0}^{\nu-1}\sum_{m=-l}^{m=l}\bkt{\nu lm}{\vec R_j}\ket{\nu lm}.
\ee
The basis is not orthonormal since the states have the overlap
%\begin{align}
%\label{eq:overlapmatrix}
%\bkt{J}{J'} &= \sum_{lm}\sum_{l'm'}\bkt{\vec R_j}{\nu lm}\bkt{\nu lm }{\nu l'm'} \bkt{\nu l'm'}{\vec R_{J'}}\nonumber\\
%&= \sum_{lm}\bkt{\vec R_j}{\nu lm}\bkt{\nu lm}{\vec R_{j'}} = \bkt{\vec R_j}{J'}.
%\end{align}
\begin{align}
\label{eq:overlapmatrix}
\bkt{J}{J'} &=
 \sum_{lm}\bkt{\vec R_j}{\nu lm}\bkt{\nu lm}{\vec R_{j'}}\\&\nonumber = \bkt{\vec R_j}{J'}.
\end{align}
 The matrix representation of $H_0$ within this basis is
\begin{align}
H_{JJ'}
&= 2\pi\sum_{\{K,L\}=1}^Ma_s(R_i)\bkt{J}{K}^{-1}\bkt{K}{L}\bkt{L}{J'}\nonumber\\
&= 2\pi a_s(R_j)\bkt{J}{J'}\label{eq:tightbinding},
\end{align}
which accounts for the requisite overlap matrix. Since we consider only a single $\nu$-manifold, we set the constant energy shift of the first term of $H_0$ to zero. For the ring geometry we will employ below, the scattering length scales out of the matrix elements since it does not depend on any relative angles between scatterers. However, to facilitate the development of Rydberg scaling laws we ignore the non-trivial $\nu$-dependence of the scattering length and set $a(R_j)\to 1$. 

%\section{Ring Rydberg Composite}

\section{Eigenspectrum of the Ring Rydberg Composite}
We now specialize to the geometry of a Ring Composite having $M$ coplanar scatterers equidistant from the Rydberg core and arranged at the azimuthal angles $\varphi_j= \frac{2\pi j}{M}$, $j=1,2,...,M$. The basis transformation used above allows us to recast the Rydberg Hamiltonian into an effective tight binding Hamiltonian operating on the states $\ket{j}$ of a 1D chain with periodic boundary conditions,
\begin{align}
\label{siteenergies}
H& = \sum_{j=1}^ME_j\ket{j}\bra{j} + \sum_{j=1}^M\sum_{j'\ne j}V_{jj'}\ket{j}\bra{j'}.
\end{align}
We distinguish the trilobite state $\ket{J}$, associated with a scatterer position $\vec R_j$, from the tight-binding state $\ket{j}$ at the same position by upper/lowercase font, respectively. Crucially, the tight-binding states $\ket{j}$  are orthogonal -- $\bkt{j}{j'} = \delta_{jj'}$ -- while the overlap matrix between trilobite states, denoted $\mathcal{O}_{jj'}= \bkt{J}{J'}$, is non-zero. The tight-binding and trilobite states at a given site are related by the transformation $\ket{J} = \sum_{j'} \left(\mathcal{O}_{jj'}\right)^{1/2}\ket{j'}$. The non-orthogonality of trilobite states leads to this non-unitary transformation matrix and the ``on-site'' energies $E_j$ and interactions $V_{jj'}$, which are obtained directly from Eq. \ref{eq:overlapmatrix}:
\begin{align}
\label{eq:hamcirc1} 
E_j   &= \frac{1}{ 4r_0^2\nu^4}\sum_{lm}F_{lm}u_{\nu l}^2(2\nu^2r_0),\\
V_{jj'}&= \frac{1}{ 4r_0^2\nu^4}\sum_{lm}F_{lm}u_{\nu l}^2(2\nu^2r_0)e^{-\frac{im2\pi}{M}(j - j')}.
\label{eq:hamcirc2} 
\end{align}
Since the Rydberg radius scales with principal quantum number as $\nu^2$ and the largest classical turning point equals $2\nu^2$, we map the ring radius onto the scaled variable $r_0\in [0,1]$ following $R=2\nu^2r_0$. The spherical harmonics evaluated in the $z=0$ plane define the factor $F_{lm}$,
\be
F_{lm} = \left\{\begin{array}{lr}\frac{(l+1/2)}{2^{2l}}\frac{(l-m)!(l+m)!}{\left[\left(\frac{l+m}{2}\right)!\left(\frac{l-m}{2}\right)!\right]^2} & l+m\text{  even}\\
0 & l+m\text{  odd.}\end{array}\right.
\ee
The $E_j$ are independent of $j$ because we have chosen an identical scattering length for all scatterers and placed them all at the same radial position; elimination of either of these constraints can be used to vary the on-site energies.

\begin{figure}[t]
\begin{center}
\includegraphics[width = 0.7\textwidth]{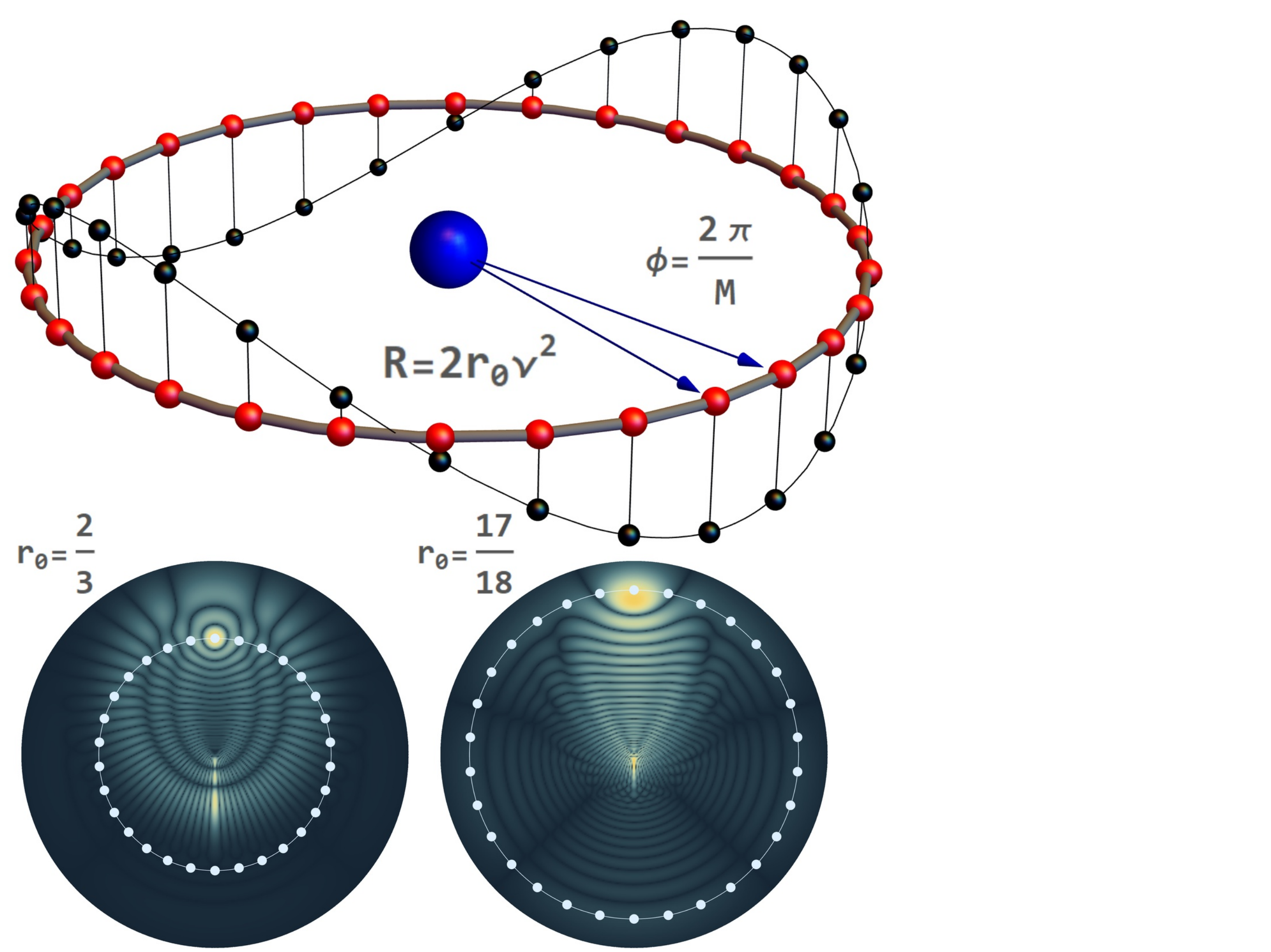}\end{center}

\caption{A diagram of the setup. The scatterers (red spheres) lie a distance $R=2r_0\nu^2$ from the Rydberg core (blue sphere); their angular separation is $2\pi/M$. The altitude of the black spheres is given by Re($c_{kj}$), the on-site trilobite amplitudes, defined in Eq. \ref{eq:estates1}. For two different $r_0$ values, the lower panels depict characteristic trilobite wave functions associated with the top site using a density plot.  The on-site energy is fixed by the amplitude of trilobite wave function at the top site, while the intersite interactions are proportional to the amplitude of this wave function at each site (blue disks).  }
 \label{fig:diagram}
\end{figure}

The $k$th eigenstate of the Ring Composite Hamiltonian $H$ is written in this site basis as 
\be
\label{eq:estates1}
\ket{\Psi_k} = \sum_jc_{kj}\ket{j};\,\,\,\,\,c_{kj} = \frac{1}{\sqrt{M}}e^{-\frac{2\pi i kj}{M}}.
\ee
The index $k=-M/2,\dots,M/2-1$ if $M$ is even and is $k=-(M-1)/2,\dots,(M-1)/2$ if $M$ is odd. 
Fig.  \ref{fig:diagram} contains a schematic of the ring geometry, using red spheres to depict the scattering sites and black spheres to illustrate the coefficients $c_{kj}$. The lower panels show how the on-site energy $E_j$ is determined by the amplitude of the trilobite wave function at that site, $\bkt{\vec R_j}{J}$, while the interaction between sites $\ket{j}$ and $\ket{j'}$, $V_{jj'}$, is determined by the off-site amplitude $\bkt{\vec R_{j'}}{J}$. 

Finally, the eigenenergies can be computed by applying the Hamiltonian to these eigenstates, giving
\begin{align}
\label{eq:evals1}
E_k(r_0)
&=\frac{M}{4r_0^2\nu^4}\sum_{lm}F_{lm}[u_{\nu l}(2\nu^2r_0)]^2\delta_{m\text{mod}M,k}.
\end{align}
Since $F_{lm}$ depends only on $|m|$, the Kronecker delta implies that $E_k = E_{-k}$.  The eigenenergy mean, $ \frac{1}{M}\sum_k E_k$, is equivalent to the eigenenergy for a single scatterer $E_1$, since
\be
\label{eq:meaninter}
\frac{1}{M}\sum_k E_k = M^{-1}\Tr H = E_1,
\ee
 using the fact that the on-site energies $E_j$ are independent of $j$ (see Eq. \ref{eq:hamcirc1}).

\subsection{Hopping: long-range interactions $V_{jj'}$ }

\begin{figure}[ht]
\begin{center}
\includegraphics[width = \columnwidth]{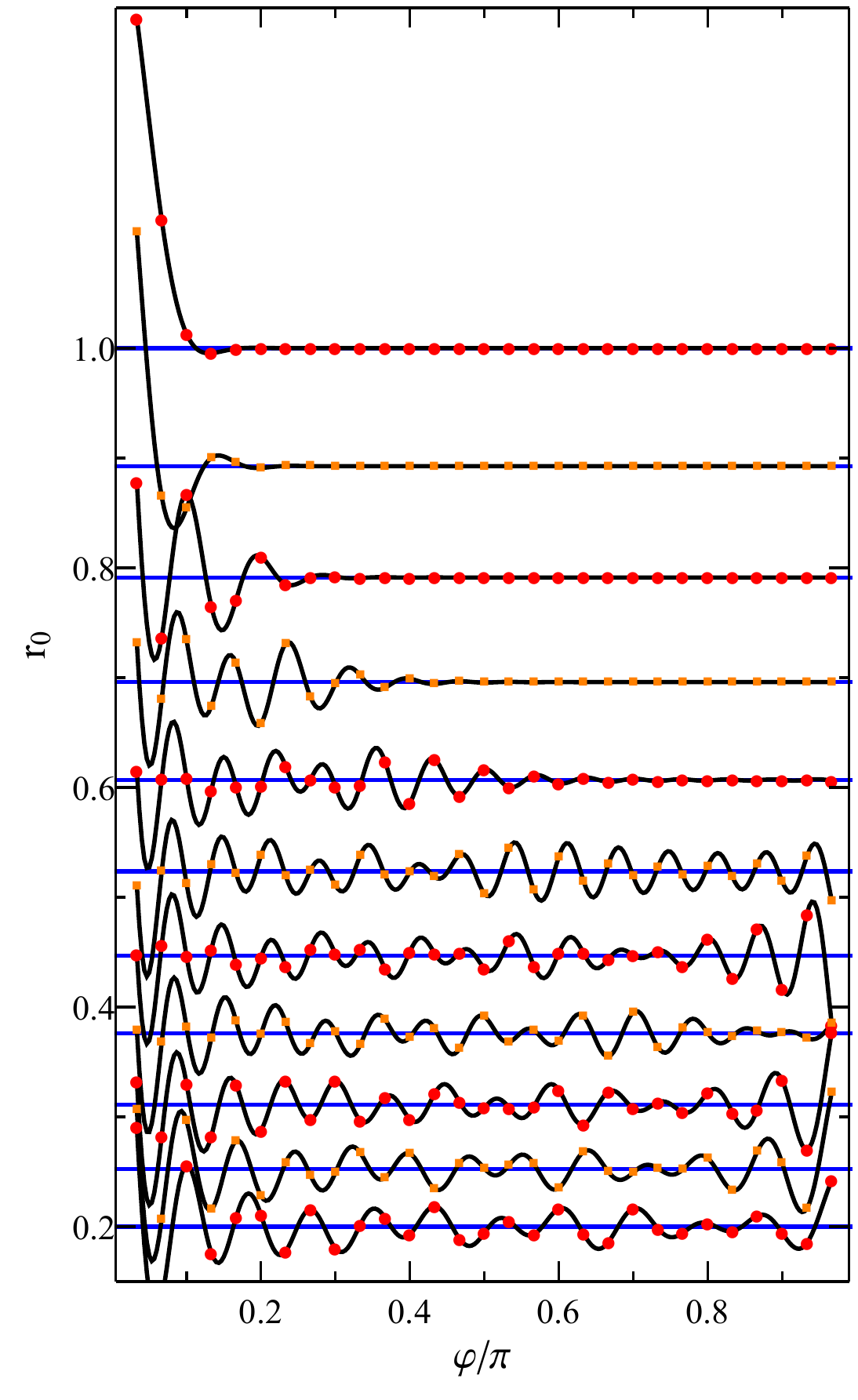}
\end{center}
\caption {The dimensionless interaction potential, $V_{1j'}/E_{1}$, is plotted as a function of $\varphi$. Only half of the possible range is shown since the interaction is symmetric about $\varphi=\pi$. The black curves show the continuum interaction ($M\to\infty$) and the alternating red circles / orange squares show the discrete values at each site for $M = 2\nu-1$; we use $\nu=30$. Each curve corresponds to a different $r_0$ value. The blue lines give the zero-interaction baseline for each $r_0$.     }
 \label{fig:interactions}
\end{figure}

The interactions $V_{jj'}$ determine the dispersion relation of the eigenvalues $E_k$ and can be tuned to realize various paradigmatic interactions as a function of ring size $r_0$.  In Fig. \ref{fig:interactions} we plot the potential $V_{jj'}$ for several characteristic values of $r_0$. At large $r_0$ the interaction is short-ranged, decaying rapidly to zero after only three sites.  As $r_0$ decreases, the interaction becomes longer-ranged, eventually spanning half of the interaction sites by $r_0\approx 0.7$, and oscillates with a larger amplitude. The oscillations, damped by the overall decay of the interaction potential, call to mind the RKKY coupling \cite{RKKY}. By $r_0 \approx 0.52$, the potential on the opposite side of the ring is non-zero, and for smaller values, such as $r_0 \approx 0.45$ or $r_0 \approx 0.31$, the largest interaction is actually between sites on opposite sides of the ring. This system therefore provides a physical realization of curious ``infinite''-range interactions. The spatial frequency of sites is typically incommensurate with the oscillation frequency of the continuum interaction, and hence the sites pseudo-randomly sample the interaction. This provides opportunities to study both uncorrelated and correlated long-range interactions. Note that the symmetry in $\pm m$ leads to real interactions despite the explicit appearance of a complex exponential in Eq. \ref{eq:hamcirc2}.

\begin{figure}[t]
\includegraphics[width = \columnwidth]{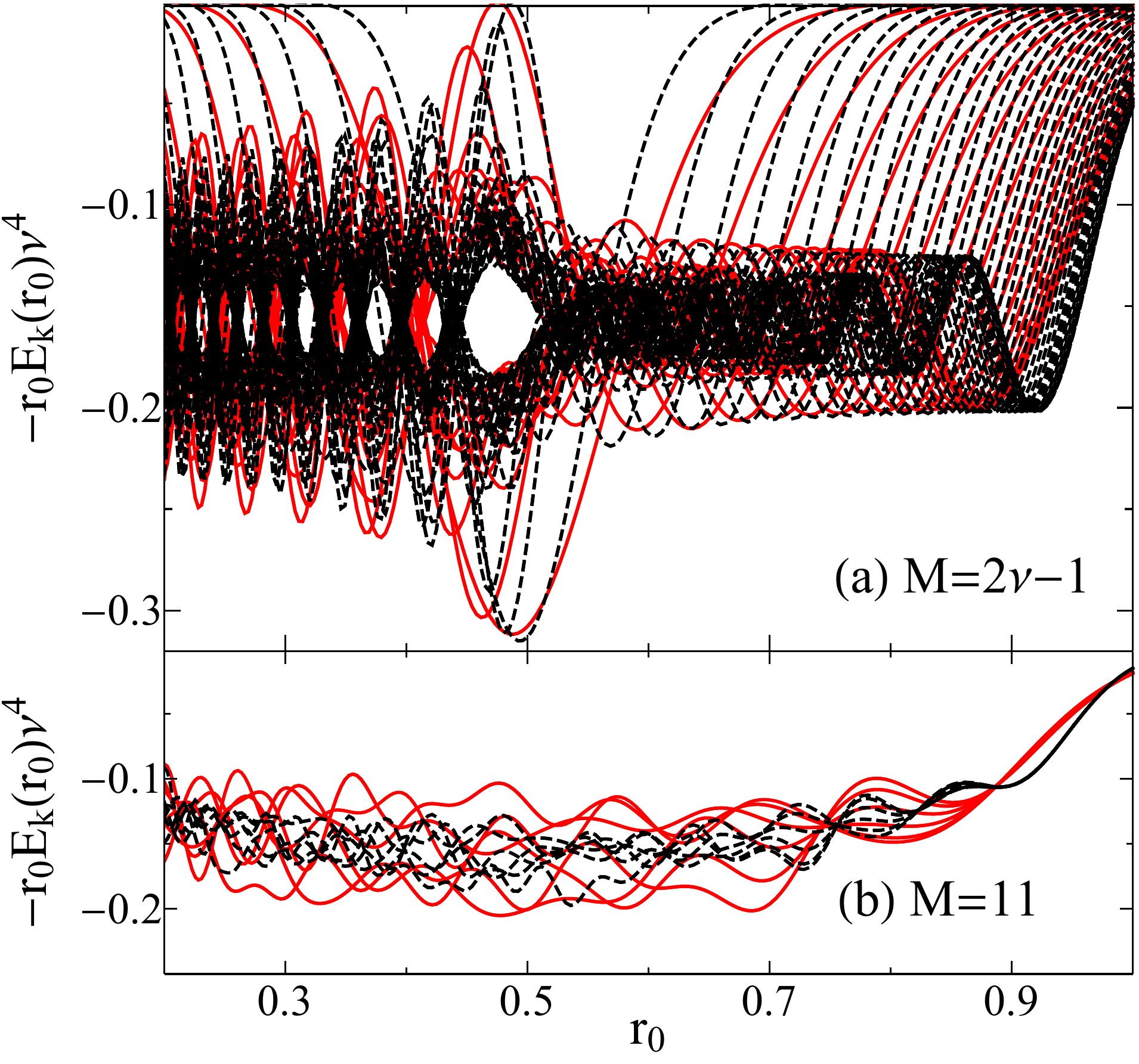}
\caption {Scaled eigenenergy curves for two different principal quantum numbers, $\nu=40$ (black, dashed) and $\nu=20$ (red), and for two different $M$ values, (a) $M = 2\nu - 1$ and (b) $M = 11$.  }
 \label{fig:potentialsnu}
\end{figure}

\section{Scaling laws: dependence on $M$, $\nu$, and $r_0$}

We now derive the scaling behavior of the eigenvalues $E_k$ as a function of $M$, $\nu$, and $r_0$. Fig. \ref{fig:potentialsnu} demonstrates the utility of these scaling laws by comparing two principal quantum numbers, $\nu=20$ and $\nu=40$, and two $M$ values, $M = 11$ and $M = 2\nu-1$. The eigenenergies have been scaled by a factor $r_0\nu^4$, and have been plotted as negative values to call to mind the original ``trilobite'' potential curves \cite{Greene2000}. After scaling, the overall amplitudes of the eigenenergy curves are constant at a coarse level in $r_0$, $M$, and $\nu$.

\begin{figure*}[t]
\includegraphics[width = \textwidth]{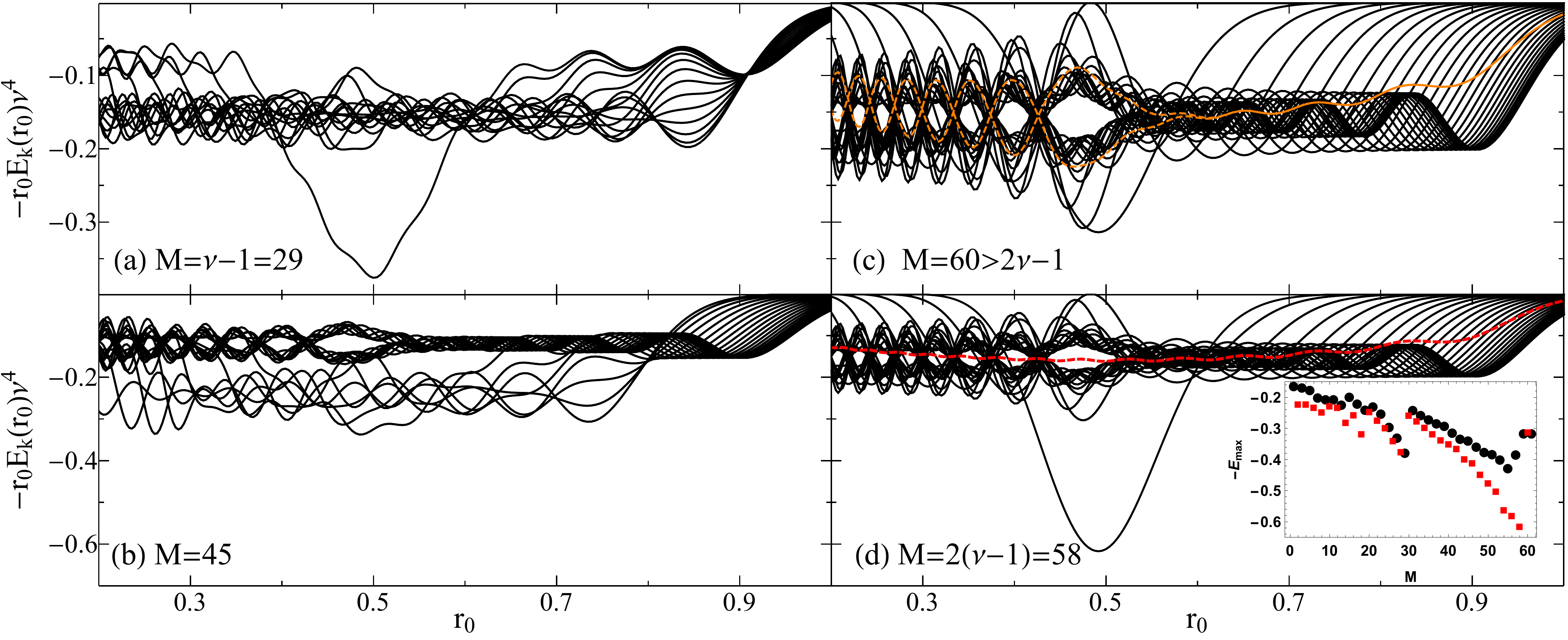}
\caption{Scaled eigenenergies for $\nu = 30$. Each panel corresponds to a different $M$. The dashed orange curve in panel (b) shows the $M=2$ result, and the dashed red curve in panel (d) shows the $M = 1$ result. Note the difference in energy axes between the top and bottom panels. The inset in (d) shows the global maxima $E_\text{max}$ as a function of $M$ using red squares for even $M$ values and black dots for odd $M$ values.} 
 \label{fig:potentialstable}
\end{figure*}

To derive the linear scaling with $r_0$, we note that the $u_{\nu l}$ radial functions oscillate in $r_0$ with an amplitude that grows approximately as $\sqrt{r_0}$. We exploit this overall dependence to find that the eigenenergies are proportional to $r_0^{-1}$. As a second general principle, we note that for each $l$, $u_{\nu l}$ decreases exponentially beyond  the $l$-dependent classical turning point, which has a maximum value of $2\nu^2$ ($r_0 = 1$) for $l=0$ states and a minimum value of $\nu^2$ ($r_0 = 0.5$) for $l = \nu-1$ states. This implies two key results: the contributions from various Rydberg states $\ket{\nu lm}$ can effectively vanish as a function of $r_0$ depending on the radial extent of that basis state, and conversely, if a given eigenstate is dominated by contributions from states with high $m$ and $l$, its eigenenergy will decrease exponentially as a function of $r_0$. The effects of these results become apparent in the discussion of Fig. \ref{fig:potentialstable} in Sec. \ref{sec:analysis}.

The $\nu^4$ scaling is clear from inspection of Eq. \ref{eq:evals1}: there is a $\nu^{-4}$ factor in front of Eq. \ref{eq:evals1} and the $u_{\nu l}$ functions inside the summation contribute an additional $(\nu^{-1})^2$. The double sum over $l$ and $m$ contributes another factor $\nu^2$, and hence the total scaling is $\propto \nu^{-4}$. The eigenenergies are approximately constant in magnitude between different $M$ values, since the factor $M \delta_{m\text{mod}M,k}$ is effectively unity. When $M>\nu-1$, the Kronecker delta function can be replaced by $\delta_{mk}$; when $M>2\nu-1$, and therefore the range of allowed $k$ values exceed $|k|=\nu-1$, this Kronecker delta condition cannot be satisfied since $|m|\le \nu-1$. It is therefore useful to explicitly define a modified form of the spectrum in Eq. \ref{eq:evals1} which is valid only for $M\ge \nu$,
\be
\label{analyticalevals}
E_k^{M>\nu}(r_0)=\Theta(\nu-|k|)\frac{M}{4r_0^2\nu^4}\sum_{l=|k|}^{\nu-1}F_{lk}[u_{\nu l}(2\nu^2r_0)]^2,
\ee
where $\Theta(x)$ is the Heaviside step function. 
%E_k(r_0)=\left\{\begin{array}{lr}\frac{M}{4r_0^2\nu^4}\sum_{l=|k|}^{\nu-1}F_{lk}[u_{\nu l}(2\nu^2r_0)]^2&\nu \le M\le2\nu-1\\
%0&M>2\nu-1
%\end{array}
%\right.
%\ee
%\be
%\label{analyticalevals}
%E_k(r_0)=\left\{\begin{array}{lr}\frac{M}{4r_0^2\nu^4}\sum_{l=|k|}^{\nu-1}F_{lk}[u_{\nu l}(2\nu^2r_0)]^2&\nu \le M\le2\nu-1\\
%0&M>2\nu-1
%\end{array}
%\right.
%\ee
The spectrum saturates when $M>2\nu-1$: the eigenenergies freeze as a function of $r_0$ and simply scale by the factor $\frac{M}{2\nu-1}$. The Ring Composite therefore has a maximum of $2\nu-1$ non-zero eigenenergies, twice the number available for a linear 1D array of scatterers \cite{PRXus}.

\section{Energy dispersion at fixed $r_0$}

 \begin{figure}[b]
\includegraphics[width =\columnwidth]{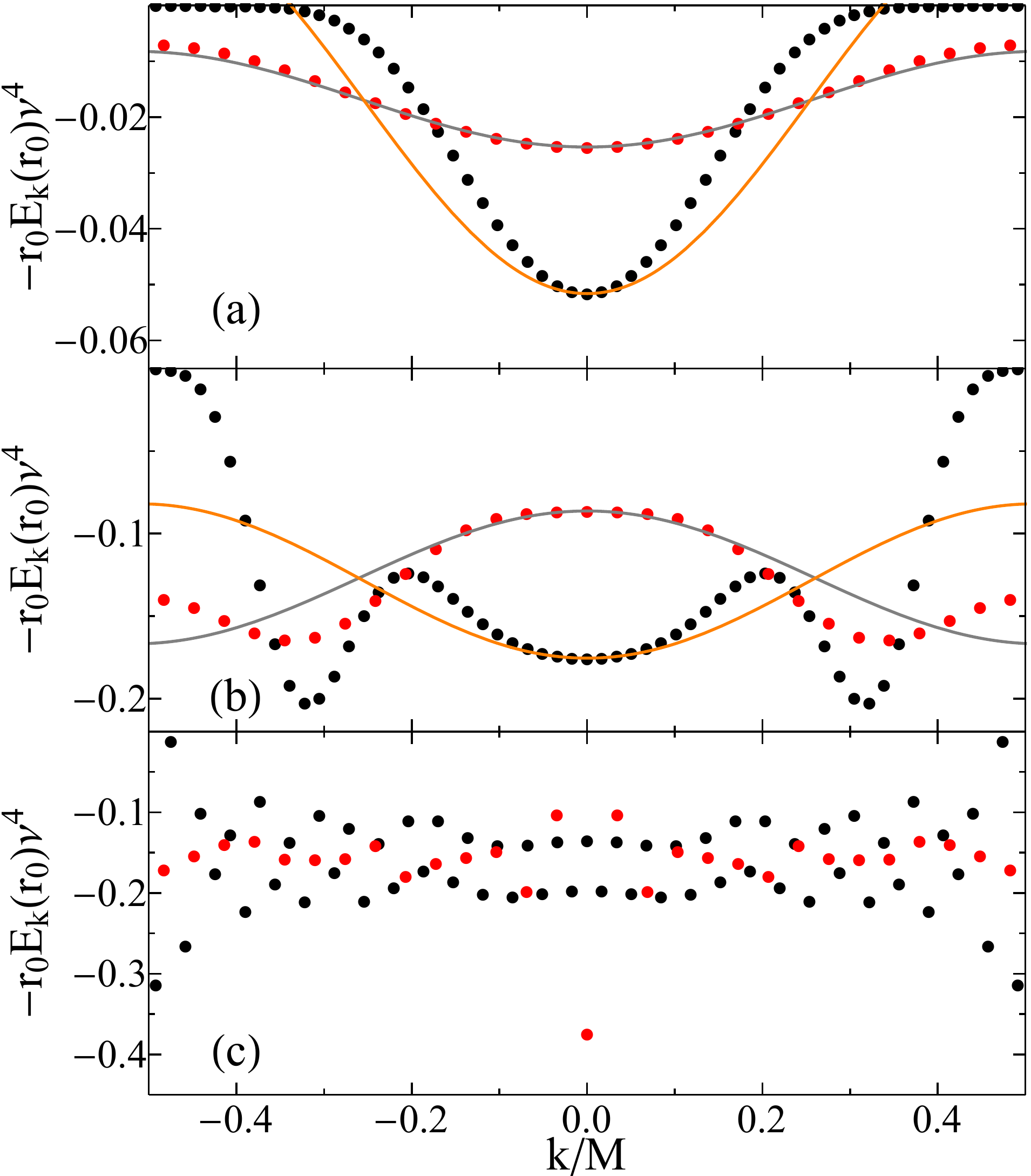}
\caption{Dispersion curves $-r_0E_k(r_0)\nu^4$ for $\nu = 30$ and three different $r_0$ values: (a) $r_0=1$,  (b) $r_0 = 0.8$, and (c) $r_0 = 0.5$. The black dots are for $M=2\nu-1$ and the red dots are for $M = \nu-1$; the orange and grey curves are the dispersion curves in the nearest-neighbor approximation $E_{nn}(k)$.  }
 \label{fig:dispersion}
\end{figure}

Before examining the variation of the eigenenergies with $r_0$, it is helpful to examine the dispersion relation $E_k$ at fixed $r_0$, since this can be related to known results from a tight-binding chain with nearest-neighbour interactions \cite{MalyshevPRB1995,FidderKnoester1991}. As $r_0\to 1$ the $V_{jj'}$ of Eq. \ref{eq:hamcirc2} become increasingly short-range, as seen in Fig. \ref{fig:interactions}. This is especially true for lower values of $M$, which have larger angular separations. We therefore approximate the $V_{jj'}$ by a nearest neighbor interaction of strength $V$ and average on-site energy $\mathcal{E}$; this model Hamiltonian has the dispersion relation $E_\text{nn}(k) =\mathcal{E}- 2V\cos\left(\frac{\pi k}{M/2+1}\right)$.  In Fig. \ref{fig:dispersion}(a) we plot $E_k(r_0 = 1)$ and $E_\text{nn}(k)$ for   $M = \nu-1$ and $M = 2\nu-1$. $\mathcal{E}$ is the average of the exact eigenvalues, while $V$ is fit to the $k=0$ energy. As surmised, the dispersion relation for the lower $M$ value is almost perfectly described by $E_\text{nn}(k)$.  The results for $M=2\nu-1$, while qualitatively similar, differ because of the presence of next-nearest neighbour terms in $V_{jj'}$; this is consistent with studies of 1D lattice systems with longer-ranged interactions \cite{MalyshevPRB1995,FidderKnoester1991}. Another difference between these two different $M$ values is number of vanishing eigenvalues when $k$ is large. The allowed $k$ values are higher for $M = 2\nu-1$ than for $M=\nu-1$, and following Eq. \ref{analyticalevals}, these high $k$ states are superpositions of high-$l$ radial Coulomb functions which are exponentially small for $r_0$ well beyond the classical turning point.

At a smaller ring radius, as shown in Fig. \ref{fig:dispersion}(b), the interaction is more long-ranged for both $M$ values. The nearest-neighbor dispersion is still accurate for low $k$ values, but the oscillatory long-range couplings lead to additional features in the dispersion relation. Finally, when $r_0$ becomes even smaller (here in Fig. \ref{fig:dispersion}(c) it is $r_0 = 0.5$) and the interaction potential is very long-range, the approximate dispersion $E_{nn}(k)$ is totally invalid. Instead, we see that the dispersion curve for $M = 2\nu-1$ splits into two bands, as also seen in Fig. \ref{fig:potentialsnu}, while the $M = \nu-1$ curve has a singularly deep minimum at $k = 0$. Both of these features will be discussed in more detail in the following section.  As a final comment, note that  -- in contrast to systems with purely short-range interactions -- the band edges do not lie, in general, at $k = 0$ or at the largest $k$ values, but vary as a function of $r_0$ due to the variation of interactions and on-site energies

\section{$M$-dependence of the eigenenergies}
\label{sec:analysis}

The scaling laws derived previously show that, on a coarse scale, the properly scaled eigenenergies are comparable across differing $\nu$, $r_0$, and $M$ values. The four panels in Fig. \ref{fig:potentialstable} confirm this: the eigenenergies for the four $M$ values shown all oscillate as a function of $r_0$, reflecting the radial oscillations of the Rydberg wave function, and the majority of the scaled energies lie in the interval $[-0.2,-0.1]$. The $M=1$ eigenenergy,  shown with the red dashed curve in Fig. \ref{fig:potentialstable}(d), is the mean eigenenergy for all cases (see Eq. \ref{eq:meaninter}). The radial envelopes of the large $k$ eigenenergy curves, which grow exponentially small at large $r_0$ as mentioned in the previous section, are apparent in panels (b), (c), and (d). This phenomenon does not occur in (a) since the allowed $k$ values are not large enough. An additional universality of these results is that, as mentioned previously, the eigenenergies cease to vary as $M$ increases beyond $M=2\nu-1$. Beyond this point the curves in Fig. \ref{fig:potentialstable}(d) are universal, only stretching by an overall factor of $M/(2\nu-1)$.

We now investigate the trends and unique features at certain $M$ values which are not scale-invariant. The first key feature is the pronounced clustering of eigenenergies around two bands oscillating out of phase with one another, seen in panels (b), (c), and (d). In fact, this clustering occurs for all even $M$ values; the simplest case is given by the $M=2$ eigenenergies shown as dashed orange curves in Fig. \ref{fig:potentialstable}(c)). When $M$ is odd the situtation is more complex. For $M<\nu$, this clustering does not occur, as exemplified in panel (a). Once $M\ge\nu$, the eigenenergies for both odd and even $M$ split into three main groupings, as in panel (b). Two of these groupings resemble the alternating bands of the $M=2$ eigenenergies, regardless of the parity of $M$. The third grouping is deeper in energy and, for even $M$, splits again into two intertwined clusters. This does not occur for odd $M$, as in panel (b). As $M$ increases further the populations of the groupings change and more states fall into the first two groups, until all eigenenergies exhibit these mirroring oscillations as seen in panels (c) and (d). These twin bands are also visible in Fig. \ref{fig:dispersion}(c). 

\begin{figure}[t]
\includegraphics[width =\columnwidth]{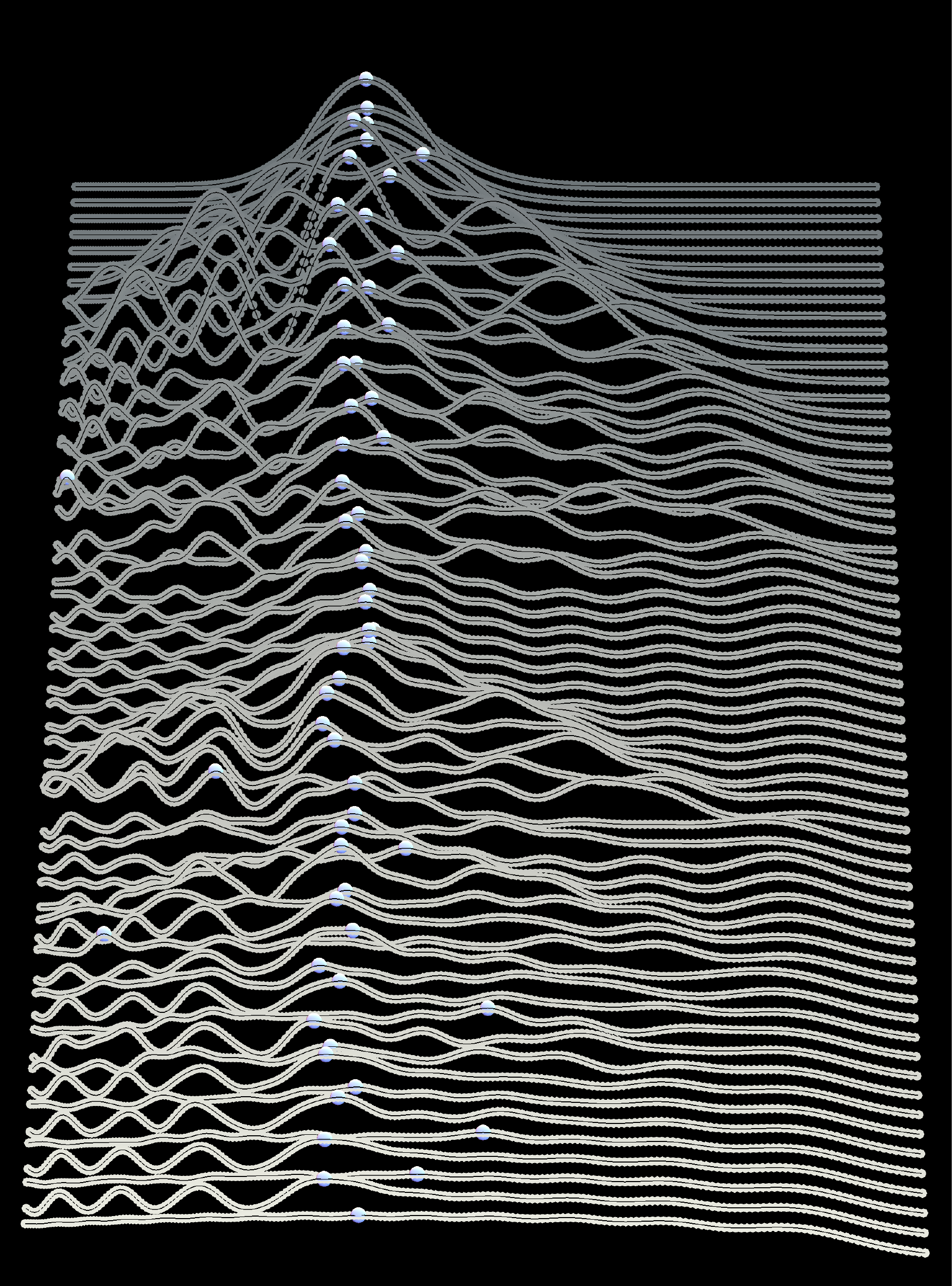}
\caption{The curves $E_{k'}(r_0)$ as a function of $r_0$, which increases from 0.2 to 1 along the horizontal axis, for each $M$, which increases from 1 to $62$ along the vertical axis. We use $\nu=30$. Local maxima, $E_\text{max}$, are marked by spheres.}
 \label{fig:potentialsfew}
\end{figure}

The second dominant feature is the existence of a remarkably deep $E_k(r_0)$ curve at specific values of $M$, visible in panels (a) and (d) as well as for $k=0$ in the $M=\nu-1$ dispersion plot in Fig. \ref{fig:dispersion}(c). We label this curve $E_{k'}(r_0)$, and its maximum value $E_\text{max}$.  In Fig. \ref{fig:potentialstable}(d) this curve is particularly pronounced, peaking at almost three times the magnitude of any other energies. The inset of \ref{fig:potentialstable}(d) shows that $E_\text{max}$ grows, although not monotonically,  as $M$ increases from $1$ to $\nu-1$.  The maxima for even $M$ values are roughly equal to or larger than the nearby odd $M$ values. For $2(\nu-1)\ge M>\nu$, $E_\text{max}$ grows monotonically when separated into distinct even or odd curves as a function of $M$. The maxima for even $M$ are much larger in amplitude and deepen at a faster rate than maxima for odd $M$. This progression is shown in greater detail in Fig. \ref{fig:potentialsfew}, where we plot $E_{k'}(r_0)$ for many $M$ values, increasing from 1 at the bottom of the figure to $61$ at the top. The radius $r_0$ ranges from $0.2$ to $1$ from left to right. The amplitude of the low-$M$ oscillations is much larger when $M$ is even than when it is odd, consistent with the splitting of the two clusters discussed above. $E_\text{max}$ is, with a few exceptions, found at ring radii close to $r_0 = 0.5$.

This behaviour is explained by considering the ability of the Rydberg wave function to adapt to the scatterer geometry, as the deepest energy shift is directly proportional to the overlap between the wave function and scatterer positions. For $M=2(\nu-1)$, where the maximum $k$ value is $|k| = \nu-1$, Eq. \ref{analyticalevals} reveals that this state involves only a contribution from the circular Rydberg state having $l = m = \nu-1$. This state has exactly $2(\nu-1)$ lobes lying in a plane and a radial wave function which is peaked at $r_0 = 0.5$, and it therefore perfectly matches the scatterer geometry. If two scatterers are removed, the ideal Rydberg basis function is the nearly circular state with $l = m = \nu-2$, which has $2(\nu-2)$ lobes around the circle and again matches the angular frequency of the scatterers perfectly. The radial wave function is not as well localized radially as before, and hence there is a corresponding decrease in the energy shift. This logic explains the steady growth in $E_\text{max}$ as $M$ increases from $\nu$ to $2(\nu-1)$, as well as the difference between odd and even $M$ values, as an odd $M$ is incommensurate with the even number of wave function lobes. However, this breaks down for the addition of just one extra scatterer to reach $M\ge 2\nu-1$: the Rydberg state, having a maximum of $2\nu-2$ angular nodes, cannot simply add another angular node to match this scatterer geometry. The much poorer overlap between the circular state and the scatterer geometry leads to a sharp reduction in the energy shift. In Fig. \ref{fig:potentialstable}(c) we can see that the deepest energy curve still has the characteristic envelope of a circular state radial function, but due to this poor overlap this energy is only slightly different from the others rather than dramatically deeper. This pronounced asymmetry in energy shift around the peak value for $M = 2(\nu-1)$ scatterers should be large enough to observe experimentally \cite{pvtcgross}.

\section{Conclusions}
We have formulated and characterized the Ring Rydberg Composite, utilizing the transformation from the usual basis of Rydberg states into the trilobite site basis to link this system to a versatile class of tight-binding lattice Hamiltonians. These possess a diverse range of interactions and dispersion relations, ranging from those described by the well-known tight-binding Hamiltonian with nearest neighbor hopping to less well-known cases involving long-range correlated or uncorrelated, RKKY-type, and even infinite-range, are possible. The Ring Composite could be coupled to an environment, driven by external fields, or change dynamically by relaxing the frozen gas assumption. The dispersion relations could be modified by introducing static or motional disorder into the scatterer locations, by coupling with additional rings of scatterers, mixing more than one atomic species together, or including spin-dependent scattering channels. 

We have focused on the theoretical promise of such a system, and here only briefly comment on its experimental realization. The recent progress in trapped Rydberg atoms and spatially designed trap arrays \cite{Birkl2019,Gross2019,Lukin2017,Browaeys2018} suggest that construction of such a composite in, for example, a tweezer array, is feasible in the future. A two-dimensional optical lattice with small lattice spacing and controllable filling could also approximate a ring geometry. Alternatively, circular neutral atom traps or lattice potentials, motivated by other theoretical proposals (e.g. \cite{PhysRevA.99.013604,amico2014superfluid}), have been created, although currently at larger sizes than desirable here 
\cite{OptFerris,FootRing,zimmermann2011high,PhysRevA.74.023617,houston2008reproducible}.

\ack{
We acknowledge funding from the DFG Priority Programme SPP 1929 (GiRyd).
MTE acknowledges support from the Max-Planck Gesellschaft via the MPI-PKS visitors program and from an Alexander von Humboldt Stiftung postdoctoral fellowship. We enjoyed many helpful discussions with A. Eisfeld. }
\section*{References}

\providecommand{\newblock}{}

%\bibliography{../../../Research/Prelim_Documents/Thesis_bib}
\end{document}